# Timing jitter in photon detection by straight superconducting nanowires: Effect of magnetic field and photon flux

Mariia Sidorova, Alexej Semenov, Heinz-Wilhelm Hübers
*DLR Institute of optical systems, Rutherfordstrasse 2, 12489 Berlin, Germany*

Artem Kuzmin, Steffen Doerner, K. Ilin, Michael Siegel
*Institut für Mikro- und Nanoelektronische Systeme (IMS), Karlsruher Institut für Technologie (KIT), Hertzstr. 16, 76187 Karlsruhe, Germany*

Ilya Charaev
*Department of Electrical Engineering and Computer Science, Massachusetts Institute of Technology, Cambridge, Massachusetts 02139, USA*

Denis Vodolazov
*Institute for Physics of microstructures, Russian Academy of Science, 603950 Nizhny Novgorod, GSP-105, Russian Federation and Physics Department, Moscow State University of Education, Moscow, Russia*

Abstract.

**We studied the effect of the external magnetic field and photon flux on timing jitter in photon detection by straight superconducting NbN nanowires. At two wavelengths 800 and 1560 nm, statistical distribution in the appearance time of the photon count exhibits Gaussian shape at small times and exponential tail at large times. The characteristic exponential time is larger for photons with smaller energy and increases with external magnetic field while variations in the Gaussian part of the distribution are less pronounced. Increasing photon flux drives the nanowire from quantum detection mode to the bolometric mode that averages out fluctuations of the total number of nonequilibrium electrons created by the photon and drastically reduces jitter. The difference between Gaussian parts of distributions for these two modes provides the measure for the electron-number fluctuations. Corresponding standard deviation increases with the photon energy. We show that the two-dimensional hot-spot detection model explains qualitatively the effect of magnetic field.**

## I. INTRODUCTION

Variance in the time delay between appearance of a photon at the optical input of a photon counter and arrival of corresponding electric signal at the pulse recorder, also known as system timing jitter, is an important property of any photon counter. For photon counters utilizing superconducting nanowires, considerable experimental [1, 2, 3, 4, 5] and theoretical [6, 7, 1] advances have been recently made in an attempt to find fundamental limits on the value of the system jitter and to reach its record values. It became clear that the measured system jitter accumulate different contributions such as variable optical delay [8] or electric noise [9, 2]. Increasing the length of nanowire introduces geometric jitter due to the position dependent travelling time of the electrical signal from the absorption site to the pulse recorder [4] In practical devices, these contributions together often exceed the intrinsic jitter contributed by the photon detection in the superconducting nanowire. The last records of the intrinsic jitter (full width of half maximum) at 1550 nm was reported to be fewer than 5 ps [3] for NbN nanowire, 20 ps for MoSi meander [5], 5 ps for NbTiN meander [10].

Ultimate, intrinsic jitter was attributed in the deterministic regime to either position-dependent growth-time of the hot-spot [1] or Fano fluctuations in the total number of non-equilibrium electrons [6] and in the probabilistic regime to the random flight-time of magnetic vortices across the nanowire [7] and to the random vortex-waiting time [1, 7].

Practical devices typically implement meander layout which consists from straight pieces of a nanowire connected by bends. It was theoretically predicted [11] and experimentally confirmed [12] that current density crowds near the inner corner of a bend. This produces an undesirable reduction of the experimental critical current. Moreover, the bends become a dominating source of dark counts and respond probabilistically to photons which are detected deterministically in straights. In our previous work we studied meandering nanowires of different sizes and were able to separate the contributions of bends and straights to the timing jitter. Although we observed the difference in jitter added by bends and by straights, the separation was not very precise.

Here we studied jitter in straight (bend-free) nanowires at the wavelengths 800 and 1560 nm corresponding to the probabilistic and deterministic detection regimes. We show that experimental statistical distribution in the time of the appearance of the photon count is best described by exponentially modified Gaussian distribution. We further show that the external magnetic field widens both the exponential and the normally distributed parts of the distribution but the rate of the increase in the exponential part is larger. We explain our finding qualitatively attributing different parts of the distribution to a sequential combination of hot-spot growth and vortex crossing and invoking spatially non-uniform hot-spot detection scheme. Comparing jitter obtained in quantum and bolometric regimes, we extract the contribution of the hot-spot growth to the jitter and show that this contribution increases with the photon energy. Experimental details are reported in Section II. Section III and IV contain experimental results and their discussion, respectively. In the Appendix A, we describe evaluation of the optical contribution to the measured jitter. Appendix B describes in details the fitting procedure and how the joint probability density for the system jitter is build up from probability densities of a series of sequential contributions.

## II. EXPERIMENT

### A. Sample preparation

We studied timing jitter in straight NbN nanowires of the lengths 40 μm (Fig.1). The nominal width of nanowires was $w = 100$ nm. Superconducting NbN films with a thickness of $d = 5$ nm were deposited on $Al_2O_3$ substrate by RF magnetron sputtering. The nanowire was drawn by the negative electron-beam lithography which results in significant improvement of the superconducting characteristics [13]. The active wire was surrounded by parallel equally spaced and electrically suspended wires of the same width (Fig. 1) in order to eliminate diffraction and to obtain the same optical coupling as for meander layout. The active wire was connected to contact pads. One of them shortened a coplanar transmission line. From the fit of the resistive transition with the form provided by superconducting fluctuation, we found the mean-field transition temperature $T_C = 12.55$ K. Transport measurements showed a critical current of $I_C = 50.2$ μA at 4.2 K and the normal square resistance 331.8 Ω/square at 25 K. The latter was slightly larger than the square resistance of the non-structured film at the same temperature $R_{SQ} = 260$ Ω/square. The measured parameters of the nanowire are similar to the parameters of meanders studied in Ref [1]. For our nanowire we found critical current density $j_C = I_C / (w\,d) = 10.4$ MA/cm$^2$ and residual resistivity at 25 K $\rho_0 = 165.9$ μΩ×cm.

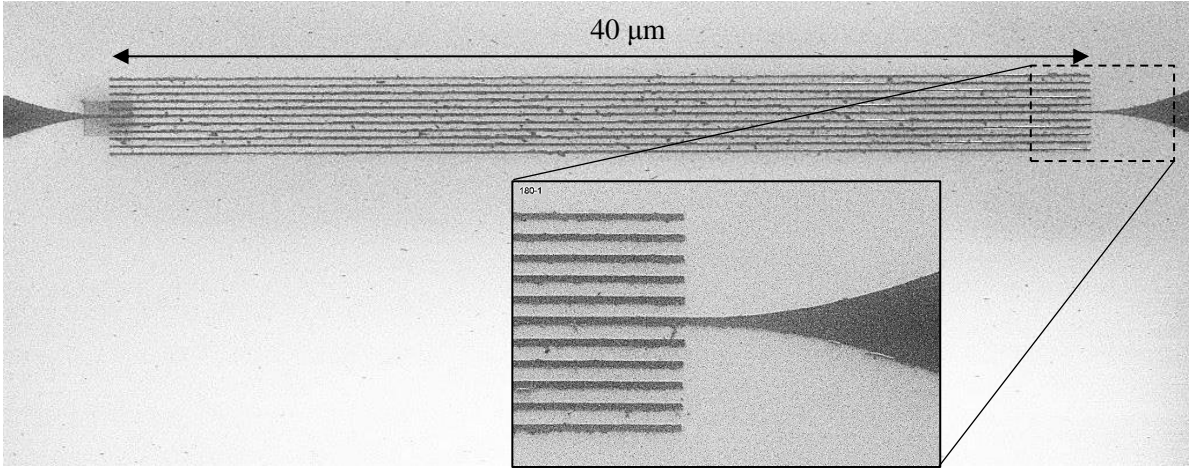

FIG. 1. Images of the nanowire obtained with a scanning electron microscope. The active wire is surrounded by parallel equally spaced and electrically suspended wires of the same width. Dark color represents NbN film.

### B. Experimental approach

We studied jitter at 4.2 K and at two wavelength 800 nm and 1560 nm. In order to minimize contributions of readout electronics and optics to the system jitter, we used pulsed lasers with sub-picosecond duration, a battery-powered DC current source, low-noise first-stage amplifier (bandpass 0.1-8 GHz, noise level 1.4 dB) and real-time oscilloscope (Keysight Infiniium X-series 93204A) with effective bandwidth for two active channels 33 GHz or alternatively the sampling scope (Keysight Infiniium DCA-X 86100D) with a bandwidth of 50 GHz. Chips with nanowires in coplanar lines were mounted on a dip-stick, with 2 meters of single-mode fibers (SMF28 for 1560nm and SM980 for 800 nm) inside which were used to deliver the light to the sample. To deliver light from the laser to the dip-stick at room temperature we used additionally 4 m and 2 m fibers (SMF28) for 1560nm and 800 nm respectively. Between the chip and the end ferrule of the inner fiber there was a distance of 15 mm so that the light spot on the substrate has 2 mm in diameter. Since the spot was much larger than the wire, we assume that it was uniformly illuminated. Magnetic field produced by a superconducting solenoid was applied perpendicularly to the substrate surface. Alternatively, the effects of the bias current and photon flux on the jitter were studied with open laser beam that allows us to eliminate completely the contribution due to variations in the flight-time of the photon through the fiber. For these measurements, specimens were mounted in a continuous-flow cryostat with optical access through a thin quartz window.

We consider the time-delay between arriving of a photon at the fiber input and arriving of corresponding electrical pulse to the recorder as a random variable. To obtain probability density function (PDF) of the arrival time, which is differently called jitter histogram or instrument response function, we measured the difference between arrival times of two voltage transients appearing at two different channels of the real-time scope. One of the transients is generated by the nanowire itself while another by a fast photodiode which was illuminated by the same laser. Triggering was done at the rising edge of the voltage transient from the nanowire. Since the arrival times of both voltage transients was associated with the 50% level of the instantaneous amplitude, the amplitude fluctuations did not contribute to the measured value. To build one PDF we accumulated arrival times of ten thousand transients from the photodiode. With the sampling scope, we used the transient from the photodiode as the trigger and accumulated points which were acquired from transients originating in the nanowire.

As the measure for system timing jitter, we used standard deviation (STD) in the measured PDF. The standard deviation for the system jitter, $\sigma_{sys}$, was obtained either numerically from the raw data or via the fitting procedure which is described in Section IV and in Appendix B. STD for the noise contribution to the system jitter was estimated as $\sigma_n = \sigma U_N \cdot \tau_r / A_{mean}$ where $A_{mean}$ is the mean transient amplitude, $\tau_r$ is duration of the rising edge of the voltage transient and $\sigma U_N$ is RMS-noise of the base line. Since the delay time and the noise are statistically independent and non-correlated variables, the noise contribution can be subtracted from the measured system jitter to obtain the noise-free system jitter

$$\sigma_s = (\sigma_{sys}^2 - \sigma_n^2)^{1/2}. \tag{1}$$

When travelling through a single-mode fiber, short optical pulse with final spectral width $\sigma_\lambda$ spreads in a wider time-interval because of the optical dispersion in the fiber material and the waveguide nature of the fiber. The former is the result of the wavelength dependence of material refractive index while the latter is due to the wavelength dependent group velocities of the propagating modes. In single-photon detection regime, this leads to different flight-times of photons through the fiber. STD of the photon flight-time through the fiber is proportional to the fiber length and the effective dispersion coefficient, which is different for different sources of pulse spreading. Optical contribution to the jitter was evaluated experimentally for both wavelengths and estimated independently from the known fiber parameters and the pulse spectrum (see Appendix A). Comparison showed that the added optical jitter is mostly due to the physical dispersion in the fiber material. We found that in our experimental setup optical jitter equals to 8 ps at 800 nm and 3 ps at 1560 nm.

### III. EXPERIMENTAL RESULTS

#### A. Critical current of the nanowire in magnetic field

In order to cross-check the quality of wire edges [14, 15, 13], we studied suppression of the critical current in our wires by external magnetic field applied perpendicular to the wire plane. The dependence $I_C(B)$ shown in Fig. 2 is symmetric and exhibits a sharp maximum at $B = 0$. At small fields $I_C$ decreases linearly with increasing $B$; the decrease slows down at fields where the wire transits from the vortex-free Meissner state to the mixed vortex state. The linear dependence of the critical current on the magnetic field is described by $I_C(B) = I_C(0)(1 - B/B^*)$, where $B^*$ is the field at which approximating straight line intersects the field axis. The transition from the Meissner state to the mixed state occurs at approximately $B^*/2$ that corresponds to the theoretical predictions of both London-Maxwell (LM) [16] and Ginzburg-Landau (GL) [15] models. For our wires we found $B^* = 734$ mT. This value is less than the values predicted by the LM model $B^*_{LM} = \eta \Phi_0/(\mu e^1 \pi \xi w)$ [16] and by the GL model $B^*_{GL} = \eta \Phi_0/(\sqrt{3} \pi \xi w)$ [15]. Here $\Phi_0$ is the flux quantum, $\xi = 4.8$ nm is the coherence length, and $w = 100$ nm the nominal width of the wire. Factor $\mu \approx 0.715$ is the ration of the current that suppresses the potential barrier for the vortex entry and the GL depairing current [16]. The coefficient $\eta$ is the ratio of the experimental critical current to the depairing critical current at $B = 0$. The density of the depairing critical current is evaluated [17] according to

$$j_{CD}(T) = KL(T) \frac{4\sqrt{\pi}(e^\gamma)^2}{21\varsigma(3)\sqrt{3}} \frac{\beta_0^2(k_B T_C)^{3/2}}{e\rho\sqrt{D\hbar}} [1 - (T/T_C)^2]^{3/2}; \quad KL(T) = 0.65 [3 - (T/T_C)^5]^{1/2}, \qquad (2)$$

were $\gamma=0.577$, $\varsigma(3)=1.202$, $e^\gamma=1.781$, $\beta_0=2.05$ is the ratio between the energy gap in NbN and $k_B T_C$, $e$ is the electron charge, and $KL(T)$ is the dirty-limit correction [18]. With the nominal width $w = 100$ nm and the thickness $d = 5$ nm of our wire, we obtained $I_{CD} = 113$ μA at T = 4.2 K and the ratio $I_C/I_{CD} \approx 0.44$ comparable to the ratios reported for straights in the meanders [1] and for straight wires [13] of the same widths.

Assuming that the wire contains superconducting core with the width $w_S < w$ and the non-superconducting edges [13] one can agree the model predictions for the value $B^*$ with the experimental data. To satisfy simultaneously the experimental normal-state resistivity of the wire one has to assign particular residual resistivity to the core, $\rho_S$, and to the non-superconducting edges $\rho_N$ and assume that the measured critical current corresponds either (GL model) to the depairing current of the core $I = j_{CD} w_S d$ or (LM model) to the current $I = \mu j_{CD} w_S d$ suppressing the barrier for vortex-entry in the core.

We took the resistivity of non-structured NbN film $\rho_S = R_{sq} d = 130$ μΩ×cm and found that the experimental $B^*$ value can be described with the pairs $w_S = 61$ nm and $\rho_N$, = 292 μΩ×cm and $w_S = 57.5$ nm and $\rho_N$, = 265 μΩ×cm in the framework of the GL and LM models, respectively. Although both approaches predict close values for the width of the superconducting core, they are noticeably less than the core-width reported for similar wires in Ref. 13. In both cases the experimental critical current amounts only at a fraction (0.57 for the GL model and 0.84 for the LM model) of the predicted critical current.

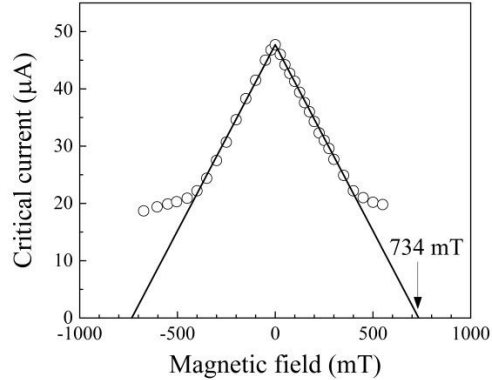

FIG. 2. Critical current of the wire in external magnetic field. Straight lines approximate the linear decrease of the current at small magnetic fields. Arrows mark the field $B^*$ discussed in the text.

### B. Fiber coupling: jitter in magnetic field

In Fig. 3 we show experimental PDFs (raw data) which were acquired for the nanowire with the length 40 μm at two wavelengths and different magnetic fields. Bias currents were $I_B = 0.63 I_C$ and $I_B = 0.75 I_C$ at the wavelength 800 nm (Fig. 4a), and 0.75 $I_C$ at the wavelength 1560 nm (Fig. 4b). At $B = 0$ mT, the PDFs for both wavelengths have almost symmetrical Gaussian shape. Increase in the magnetic field results in the appearance of the exponential tail at larger arrival times while at small arrival times PDFs retain Gaussian shape. The PDFs resemble those reported in Ref. 1 and can be best fitted with the exponentially modified Gaussian functions (Appendix B).

Solid lines in Fig. 3 show the best fits which include different contributions (optics, electrical noise) to the system jitter. We will discuss the fitting procedure in Section IV and in Appendix B. At both wavelength, the slope of the exponential tail decreases with the field that corresponds to an increase of the characteristic exponential time. At 1560 nm, STD of the Gaussian part slightly grows with the field while at 800 nm almost no changes occur. Electrical noise was found to contribute $\sigma_n$ = 4.92 ps at $I_B = 0.63\, I_C$ and $\sigma_n$ = 4.6 ps at $I_B = 0.75\, I_C$. Subtracting the noise contributions from the numerically computed STDs (Eq. 1), we obtained noise-free system jitter shown in Fig. 4. It is clearly seen that jitter increases with the magnetic field. The plateau at small fields for the wavelength 800 nm is caused by relatively large optical contribution to the jitter (8 ps).

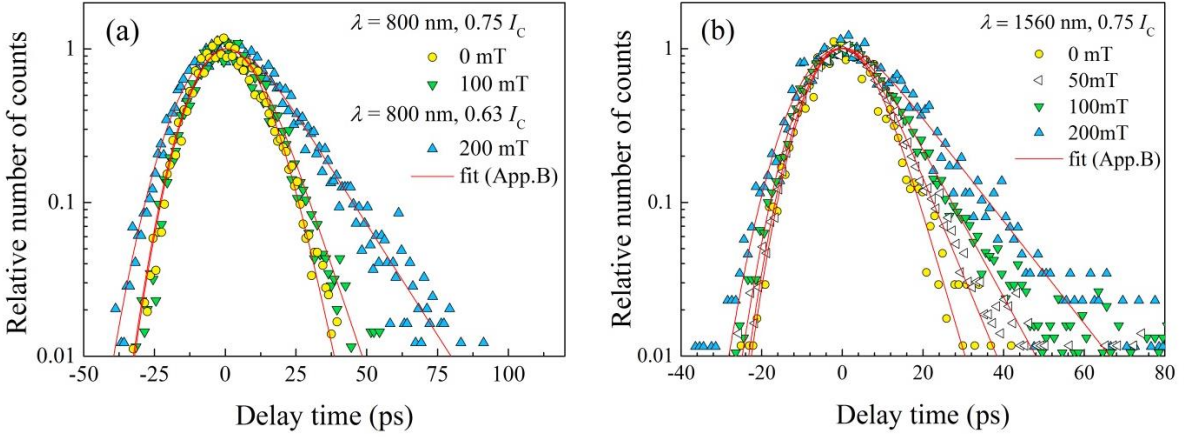

FIG. 3. (Color online) Raw PDFs in different magnetic fields. Maxima of PDFs were normalized to one and shifted to zero delays. (a) Data for the wavelength 800 nm were obtained at currents 0.63 $I_C$ and 0.75 $I_C$ and magnetic fields 0, 100, and 200 mT. (b) Data for the wavelength 1560 nm were obtained at 0.75 $I_C$ and magnetic fields 0, 50, 100, 200 mT. Solid lines show best fits.

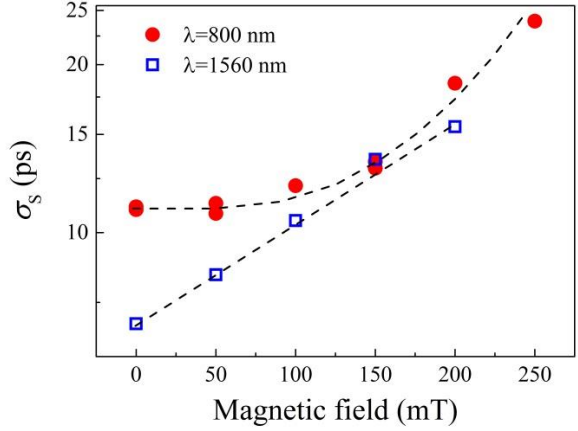

FIG. 4. (Color online) Noise free system jitter at different magnetic fields for two wavelengths 800 and 1560 nm. Standard deviations were computed numerically for each raw PDF and the noise contributions were subtracted according to Eq. 1. Dashed lines are to guide the eyes.

## C. Free-space optical coupling: different currents and photon-fluxes

Free-space coupling of light to the nanowire eliminates optical contribution to the jitter and increases accuracy of extracting components of the intrinsic jitter. Fig. 5 shows PDFs (raw data) obtained with free-space coupling at two wavelengths for different photon fluxes. It is clearly seen that the system jitter (numeric STD) obtained for both wavelengths 800 nm ($\sigma_{sys}$ = 6.7 ps) and 1560 nm ($\sigma_{sys}$ = 9.2 ps) at relatively small photon flux is noticeably larger than the system jitter $\sigma_{sys}$ = 2.2 ps at relatively large photon flux. It is also worth to note that the PDF at large photon flux is purely Gaussian, it does not contain the exponential tail typical for PDFs at small fluxes. Fig. 5b shows noise-free system jitter for two wavelengths at different currents. The jitter was measured at relatively small photon flux. Variations of the jitter with the current at 800 nm and 1560 nm qualitatively correspond to the results reported in Ref. 1 for the deterministic and probabilistic detection scenarios, respectively. At large photon fluxes, variation in the standard deviation with the current is beyond the accuracy of our measurements.

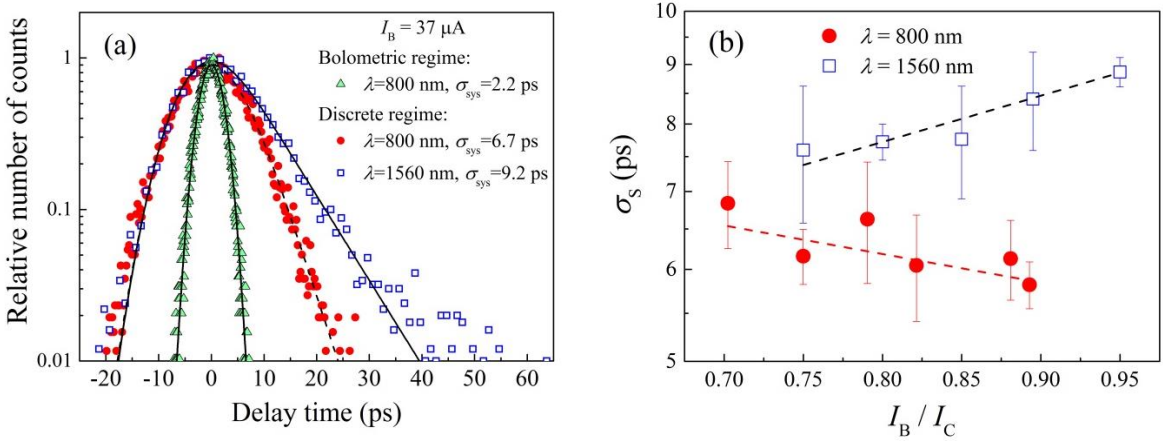

FIG. 5. (a) Raw PDFs measured via free-space light coupling for two wavelengths with two different photon fluxes at a fixed current of 37 μA: at wavelengths 800 nm (circles) and 1560 nm (squares) with relatively small flux and at the wavelength 800 nm (triangles) with relatively large flux. Solid curves show best fits described in Section IV. Legend shows system jitter (numeric STD) for corresponding PDFs. (b) Noise-free system jitter at different currents obtained with Eq. 1.

We did not define the exact values of the photon flux incident on the nanowire. Instead, we measured amplitudes of the signal transients at different fluxes but constant bias current as well as current dependences of the photon-count rate (PCR) at different photon fluxes. The data are presented in Fig. 6. At a fixed current, amplitude of the transient remains constant up to certain value of the photon flux and then increases with the further increase of the flux (Fig. 6a). At large flux values, the amplitude saturates as it is seen in the inset in Fig. 6a on the linear flux-scale. Current dependence of the photon-count rate changes drastically with the increase in the photon flux (Fig. 6b). At relative fluxes less than 0.01 (the lower curve), the dependence has the shape typical for single-photon detection. As it is seen in the inset in Fig. 6b, PCR at large currents grows linearly with the flux which additionally justifies single-photon detection regime. At large fluxes, count rate does not change with the current (upper curve in Fig. 6b) and equals the rate of laser pulses 80 MHz. This rate is smaller than the maximal count rate of the nanowire (500 MHz) which is defined by the dead time (2 ns) associated with photon count. Saturation in PCR is typical for multi-photon process with the large number of photons absorbed simultaneously. Since the incoming photons are evenly distributed over the surface of the nanowire, we conclude that the response of the nanowire is not discrete any more but rather becomes bolometric. The PDFs shown in Fig. 5 were obtained at relative flux values 0.001 and 10, i.e. in the quantum and bolometric detection regimes. This supposition is further supported by the flux dependence of the transient amplitude and its saturation at large fluxes (inset in Fig. 6a). The best fit shown in the inset was obtained with the expression $f(x) = a\, b\, x/(z0 + b\, x)$ where $z0 = 50\,\Omega$ represents the input impedance of the transmission line, $x$ is the relative photon flux, and $a$ and $b$ are adjustable parameters. The best-fit values of the parameters $a$ and $b$ are 0.5 and 31, respectively. The letter corresponds to a change of 310 Ω in the resistance of the nanowire under effect of the laser pulse for $x = 10$. Extrapolating to the discrete regime, with the kinetic inductance of our wire 6 nH and the flux-independent rise time of the transient 70 ps we arrive at the resistance of the normal domain in excess of 400 Ω that corresponds to the value estimates in the framework of the electro-thermal model [19].

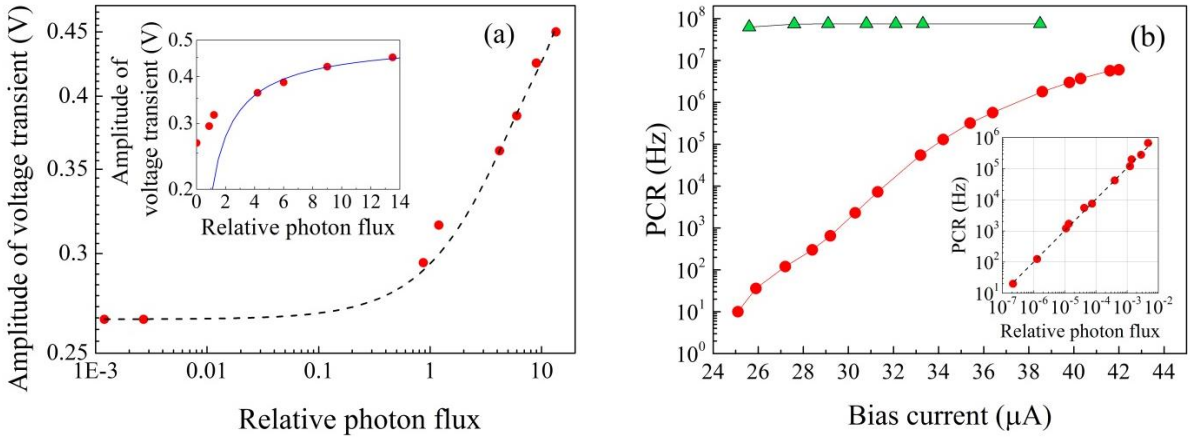

FIG. 6. (a) Amplitude of the voltage transient at 800 nm as function of the photon flux. Dashed line is a guide to the eye. Inset shows the same data on the linear flux-scale. Solid line in the inset is the best fit described in Section III C. (b) PCR as function of bias current at 800 nm and two different relative photon fluxes 0.001, and 10 for curves from the bottom to the top. Solid lines are guides to the eye. Inset shows dependence of PCR on the photon flux in double logarithmic scale. Dashed straight line shows the best linear fit.

## IV. DISCUSSION

### A. System jitter at large and small photon fluxes

Bolometric, multiphoton detection regime allows one effectively eliminate contribution of Fano fluctuations to the measured jitter. Indeed, the difference between the times of the hot-spot growth or, equivalently, the difference between energies released at many different absorption sites is averaged out when the number of sites is sufficiently large. The same is true for all jitter components associated with the start-times of vortex jumps and vortex flight-times across the wire. Additionally, this regime eliminates geometric contribution to the jitter since absorption sites are evenly distributed over the nanowire. In such regime measured jitter includes instrumental contribution (laser, oscilloscope, cables) and the contribution of electrical noise. Since both are statistically independent, the square of the measured STD is the sum of their squared STDs (Eq. 1). Raw PDF in the bolometric regime is shown in Fig. 7b (green triangles). It was obtained at 800 nm in free-space and has Gaussian shape with the standard deviation 2.2 ps. With the noise contribution of 1.2 ps this results in the instrumental contribution $\sigma_{ins}$ = 1.84 ps. Two other raw PDFs were obtained with free-space coupling in the single-photon detection regime at wavelengths 800 nm (red circles) and 1560 nm (blue squares) with system standard deviations 6.7 ps and 9.3 ps and noise contributions 2 ps and 3.95 ps, respectively. Taking the difference between noise-free system jitters in the discrete, single-photon and in the bolometric detection regimes, we obtained remaining jitters 6.1 ps at 800 nm and 8.2 ps at 1560 nm. Each of these values includes geometric contribution, and the contribution which stems from the detection process itself and which we call local jitter.

Let us estimate geometric contribution to the measured jitter. When a nanowire acts as the central line of a shortened portion of a coplanar transmission line, this contribution may appear due to variations of the propagation times of the current steps from the detection site in the nanowire to the output of the shortened portion. In the ideal case, the midtime between arrivals of two current steps (one travelling via the wire and another − via the ground plane) at the output does not depend on the position of the absorption site. It has been shown that the propagation velocity of the current step in the superconducting nanowire $v \approx 12$ μm/ps [20] which is more than one order of magnitude less than the propagation velocity in the transmission line [1]. For the nanowire with the length $l = 40$ um, the maximum difference between arrival times of two steps is $l/v \approx 3.3$ ps. This is more than one order of magnitude smaller than the time resolution ≈ 70 ps of our read-out (amplifiers, cables, scope) set by the effective upper band-pass frequency. Although the readout does not resolve two current steps, the arrival time at a fixed discriminator level should not suffer any jitter. However, the geometric jitter in the layout with the shortend transmission line may appear if the dispersion and losses are different in the nanowire and in the ground plane. Effective differential propagation velocity for our line amounts at 70 μm/ps [1] that results in $\sigma_{geom}$<0.6 ps. This value is comparable to the accuracy of our measurements. Therefore, we neglected geometric jitter in the following consideration. In order to separate contributions from Fano fluctuations and vortex jumps to the intrinsic jitter, we introduce the fitting procedure described in the next Section.

### B. Components of the local jitter

Following the formalism of our previous study [1], we represent each raw PDF as the joint PDF of a series of sequential events which are absorption of a photon in the nanowire (i), emergence of the hot-spot (ii), start of the vortex crossing (iii), and arrival of the voltage transient at the input of the first amplifier (iv). The noise of the first amplifier and the contribution of the instrument enhance the measured jitter and is considered as the last event in the series (v). Each particular event except the amplifier noise is associated with its own time-delay which is considered as particular random variable. They are flight-time of a photon through the fiber (i), growth time of the hot-spot (ii), time-delay between emergence of the hot-spot and the start of the vortex-crossing (iii), and the travelling time of the voltage transient (iv). All these variables are statistically independent but variables affiliated with adjacent events are connected via conditional probability, i.e. the later event occurs only if and after the earlier has occurred. The total delay-time is a sum of delays associated with particular events. Conditional joint PDF for two sequential events with particular PDFs $f_1(t)$ and $f_2(t)$ is given by [21]

$$f(t) = \int_{-\infty}^{t} f_1(u) f_2(t-u) du . \qquad (3)$$

Conditional probability also implies that $f_2(t) = 0$ at $t < 0$. The joint conditional PDF for a series of sequential events is a multiple integral of the type shown above which includes particular PDFs of all events in the sequence. The jitter added by the amplifier noise as well as the instrumental contributions (Section IV A) are statistically independent. They are not connected with the previous events by conditional probability (non-correlated). For a sequence of two non-correlated events, the joint PDF is also given by Eq. (3) with the upper integration limit set to plus infinity. Details of the fitting procedure and the way of building up the joint PDF are described in the Appendix B. There we also show that the system jitter can be presented as

$$\sigma_{sys} = (\sigma_{opt}^2 + \sigma_{int}^2 + \sigma_n^2 + \sigma_{ins}^2)^{1/2}, \qquad (4)$$

where $\sigma_{opt}$, $\sigma_{int}$, $\sigma_n$ and $\sigma_{ins}$ are standard deviations of particular PDFs affiliated with the optical delay (i), delay in the appearance of the photon count (ii-iv), electrical noise (v) and the instrumental contribution. In its turn, the jitter inherent in the appearance of the count, the intrinsic jitter, $\sigma_{int}$, can be presented as

$$\sigma_{int} = (\sigma_{loc}^2 + \sigma_{geom}^2)^{1/2}, \qquad (5)$$

where $\sigma_{loc}$ and $\sigma_{geom}$ are STDs for the local and geometric components of the intrinsic jitter associated with the detection process at the absorption site (ii, iii) and with the propagation of the signal transient (iv), respectively.

As discussed above we neglect the geometric jitter that reduces (5) to the identity $\sigma_{int} = \sigma_{loc}$. Since spectra of our laser pulses have Gaussian shapes, we assume that the PDF of the optical jitter is also Gaussian. The PDF of the noise contribution was measured directly and was proved to have Gaussian shape. Relying to the fact that the characteristic time $\tau$ of the exponential tail is not modified by joining particular PDFs with Gaussian shape (Appendix B), we further suppose that the PDF of the local jitter represents exponentially modified Gaussian distribution $g(t,\sigma,\tau)$ (Appendix B, (B3)). This particular PDF is the joint PDF of two sequential events (ii and iii). Its form steams from the supposition already discussed in Ref. 1 that the PDF of start times of vortex crossing (iii) represents exponential distribution with the characteristic time $\tau$ while the PDF of the growth-time of the hot-spot is most likely controlled by Fano fluctuations and is presented by a Gaussian PDF with the standard deviation $\sigma$. It is known (we also prove that in Appendix B) that the standard deviation of an exponentially modified Gausssian distribution equals $(\sigma^2 + \tau^2)^{1/2}$ that leads to the identity $\sigma_{loc} = (\sigma^2 + \tau^2)^{1/2}$.

In the fitting procedure, we use $\tau$ and $\sigma$ as two independent fit parameters to best fit the measured system PDFs. Figs 7 and 8 shows field and current dependences of the best fit values of $\tau$ and $\sigma$.

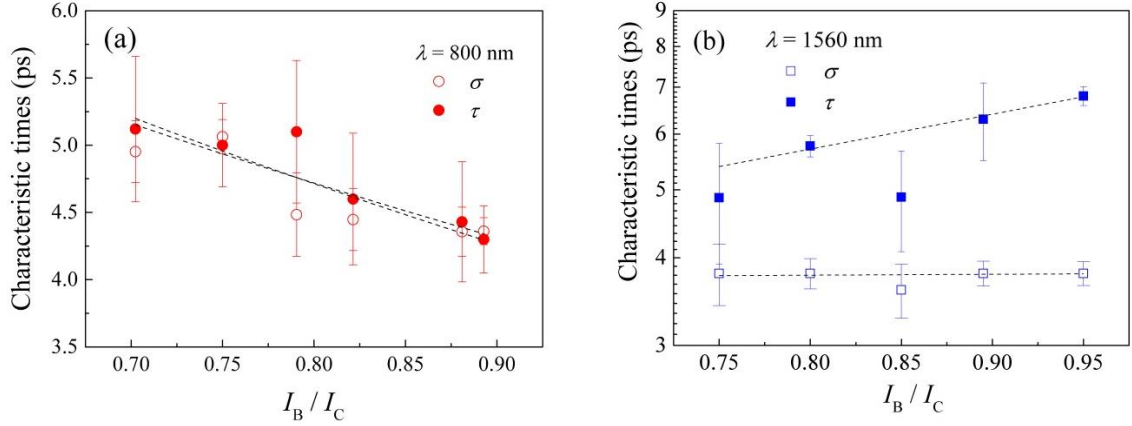

FIG. 7. Characteristic times $\tau$ and $\sigma$ of the exponentially modified Gaussian distribution associated with the intrinsic jitter vs relative bias current for $B = 0$ and two wavelength (a) 800 nm and (b) 1560 nm. Values $\tau$ and $\sigma$ are obtained with the fitting procedure described in the text.

At 800 nm (Fig. 7a) $\tau$ and $\sigma$ are almost equal and both decrease with the current while at 1560 nm (Fig. 7b) $\sigma$ is almost current independent and $\tau$ increases with the current. At currents $I_B \geq 0.9\ I_C$ typical for applications of nanowire detectors, $\tau$ for the wavelength 1560 nm is twice as large as for the wavelengths 800 nm providing the major contribution to the local jitter at the longer wavelength. Subtracting instrumental contribution to the jitter (1.84 ps, Section IV A) from $\sigma$ values, we obtain the contribution to the jitter, provided by Fano fluctuations. For $I_B = 37$ μA (Fig. 5) corresponding to $I_B = 0.88\ I_C$ we arrive at the values 4.3 ps and 3.1 ps for 800 nm and 1560 nm, respectively. The ratio between these values is very close to the square root of two. Exactly this ratio is expected for Fano fluctuations [22] which provide standard deviation in the number of excited electrons $(\varepsilon F (h\nu)/\Delta)^{1/2}$, where $h\nu$ is the energy of incident photon, $\Delta$ the superconducting energy gap, $F = 0.2 - 0.3$ the Fano factor, and $\varepsilon \approx 0.15$ the quantum yield. Hence, the standard deviation in the energy delivered to the hot-spot should scale as square root of the photon energy. A microscopic two-dimensional model of the hot-spot is required in order to translate STD in the number of excited electrons to the STD in the growth-time of the hot-spot.

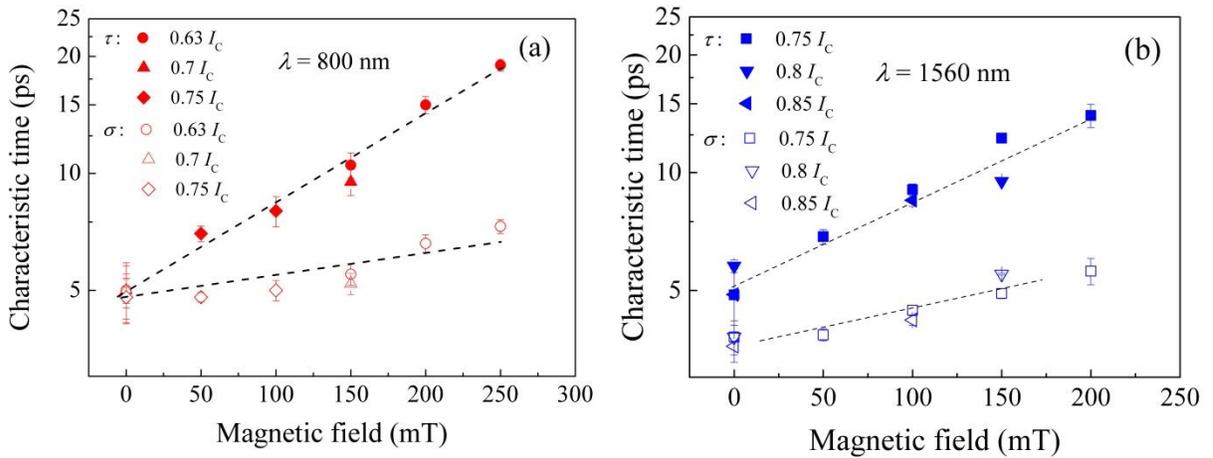

FIG. 8 (Color online) Characteristic times, $\tau$ (closed symbols) and $\sigma$ (open symbols), of exponentially modified Gaussian PDFs associated with internal detection process vs magnetic field. (a) For the wavelength 800 nm at bias currents 0.63 $I_C$, 0.7 $I_C$, and 0.75 $I_C$. (b) For the wavelength 1560 nm at bias currents 0.75 $I_C$, 0.8 $I_C$, and 0.85 $I_C$.

Data presented in Fig. 8 show that at both wavelengths $\tau$ and $\sigma$ grow with the magnetic field. Extracting known optical jitter (Appendix A) via the fitting procedure (Appendix B) allowed us to reconstruct the dependence of the local jitter for 800 nm at small magnetic fields. We found that the full variation of $\tau$ in the field range from zero to 250 mT is larger for 800 nm than for 1560 nm. Increase of the local jitter in magnetic field is inconsistent with the one-dimensional model of Fano fluctuations [6] where the field suppresses the detection current. Below we describe our finding qualitatively in the framework of the microscopic hot-spot model and the position dependent detection current.

To calculate delay time and its dependence on the magnetic field we use the two-temperature model coupled with the modified time-dependent Ginzburg-Landau (TDGL) equation [23]. We did not expect that it gives us quantitatively correct results because it is based on the assumption that electrons and phonons are instantly in the internal thermal equilibrium. This means that their distribution functions represent at any time Fermi-Dirac and Bose-Einstein distributions, respectively, with two different effective temperatures which are both larger than the bath temperature. The limited validity of this approximation was thoroughly discussed in Ref. 23. However, this is the simplest model which is capable to explain qualitatively temporal response of a superconducting wire to absorption of the photon. In spite of relative simplicity, this model takes into account important physical effects - finite relaxation time of the superconducting order parameter and its spatial variation in the hot-spot region, heating of electrons due to Joule dissipation, and growing of the normal domain. In our model vortices appear naturally at the place where superconductivity is sufficiently weakened [24]. It does not require any entry-barrier for vortices as it is the case e.g. in the London model [16, 7, 25]. The validity of TDGL equation at relatively low temperatures is justified since inside the hot region, where the dynamics of the order parameter is most pronounced, local temperature is close to $T_C$. Moreover, we checked that at $T \leq T_C$ our model gives practically the same results as at low temperatures, which further supports the usage of TDGL equation at low temperatures. We simulate voltage response of an NbN nanowire with typical material parameters also used in Ref. 23. These parameters are close to the parameters the nanowires reported here. We set the escape time for phonons at $\tau_{esc} = 0.05\ \tau_0$ (for definition of $\tau_0$ see Eq. 6 in Ref. 23) and use convenient in numerical calculations scales for distances $\xi_C = [\hbar D/(k_B T_C)]^{1/2}$ and for magnetic fields $B_C = \Phi_0/2\pi\xi_C^2$. Time is measured in units $\tau_C = \hbar/(k_B T_C)$. Computation was performed for the strip width $w = 20\ \xi_C$, ambient temperature $T = 0.5\ T_C$ and the bias current $I_B = 0.55\ I_{CD}$. The effect of the absorbed photon is modelled as instantaneous heating of electrons and phonons at $t = 0$ in the finite area $2\xi_C \times 2\xi_C$.

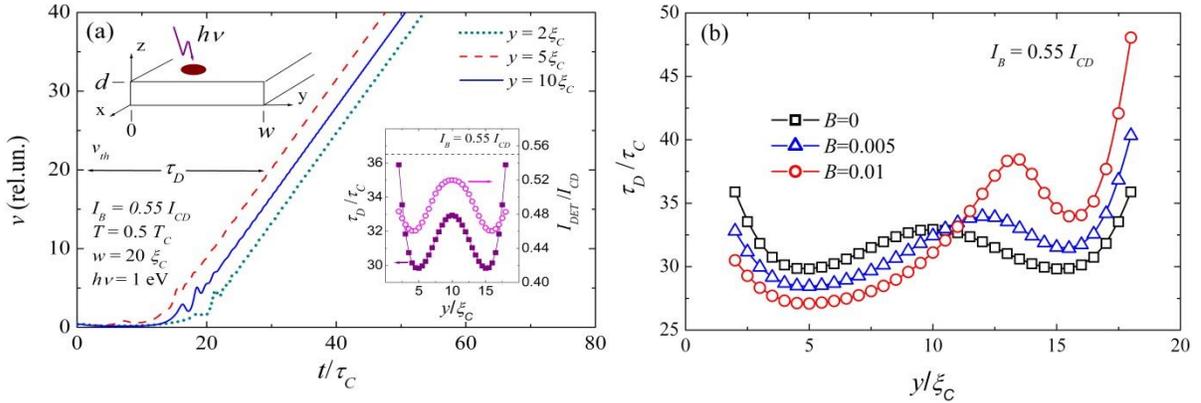

FIG. 9. (a) Time dependent voltage response of the superconducting strip in units of $k_B T_C/(2 e)$, $e$ is the electron charge, after absorption of the photon at $t = 0$ at different distances $y$ from the strip edge. The inset in the upper left corner presents the geometry. The distances are specified in the legend (upper right corner) in units of $\xi_C$. Another legend (lover left corner) specifies bias current, bath temperature, the strip width and the photon energy. Time is given in units $\tau_C$. The delay time $\tau_D$ is defined at the threshold level $v_{th} = 20$. The inset in the lover right corner shows the detection current in units of the depairing current and the relative delay time as function of the position of the hot-spot. (b) Relative delay time as function of the hot-spot position for three different magnetic fields. Fields are specified in the legend in units of $B_C$. Lines are guides to the eyes.

Fig. 9a shows the dynamics of the voltage response after absorption of the photon with energy of 1 eV at three different locations across the strip. Because of the linear increase of the voltage with time, the threshold level $v_{th} = 20$, which we use to define the delay time $\tau_D$ does not affect the variance in $\tau_D$ (timing jitter). However, for the reconstruction of the respective PDF it is important that the smallest delay time $\approx 15\ \tau_C$ is larger than the spread in $\tau_D$ values. For particular location of the hot-spot, the detection current is the bias current at which the photon absorbed in this location is detected deterministically. Inset in Fig. 9a shows position dependences of the detection current $I_{DET}(y)$ similar to the results reported earlier [23, 26]. Position dependence of $\tau_D$, which is seen in the inset, steams from the position-dependent detection current and is formed by monotonous decrease of $\tau_D$ with the increase in the ratio $I_B/I_{DET}(y)$. Indeed, our problem with delay time is physically equivalent to the well know problem with time delay in the appearance of the voltage response after instant switching on the supercritical current in a superconducting strip [27-31]. In our case, $I_{DET}(y)$ plays the role of the critical current. However, contrary to the above problem, $\tau_D$ does not diverge for $I_B > I_{DET}(y)$ because of the finite lifetime of the hot-spot. It was found [7-31] that $\tau_D$ monotonically decreases with the current and so does local $\tau_D(y)$ in our problem. At fixed applied current, $\tau_D$ is smaller at those locations in the strip where the ratio $I_B/I_{DET}(y)$ is larger. If external magnetic field is small enough (field interval in Fig. 2 where $I_C(B)$ decreases linearly) it only induces the screening current in the strip which remains in the vortex free regime. In the absence of magnetic field, density of the applied current is uniform across the strip. With the field, the density of the local current is the sum of the density of the applied current and the density of the screening current. This tilts the uniform distribution. Consequently, the dependence $I_{DET}(y)$, which is symmetric at $B = 0$ with respect to $y = w/2$ (inset in Fig. 9a), becomes asymmetric. $I_{DET}$ increases at the edge, where the density of the local current decreases, and decreases at the opposite edge [23, 26, 32]. The change in $I_{DET}$ cases respective change in $\tau_D$. Fig. 9b shows the $\tau_D(y)$ dependence in two different magnetic fields and in the absence of the field. It is clearly seen that the variance (spread) in $\tau_D$ increases with the field. Obviously, the sign of this effect does not depend on specific choice of parameters. The effect steams from the finite relaxation time of the order parameter and its dependence on the local current density which changes in the magnetic field.

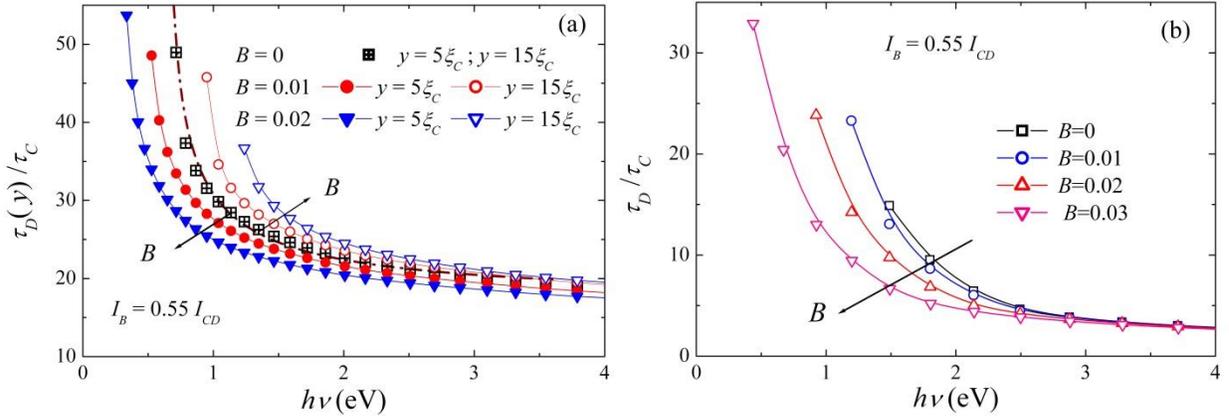

Fig. 10. Relative delay time as function of the photon energy at different magnetic fields for the bias current $0.55\ I_{CD}$ and $T = 0.5\ T_C$. Fields are specified in insets in units of $B_C$. Arrows show the direction of changes when the field increases. Solid lines are to guide the eyes. (a) Delay time for different locations of the hot-spot in the framework of the hot-spot model. Distances from the strip edge are specified in units of $\xi_C$. Dash-dotted line is the fit described in the text. (b) Delay time in the framework of the uniform hot-belt model.

It has been shown that $I_{DET}(y)$ shifts to smaller currents and flattens when the photon energy increases [23, 26, 32]. Hence, $\tau_D$ must depend on the photon energy too. Fig. 10 shows computed dependences of the delay time on the photon energy in the framework of our two-dimensional hot-spot (a) and the uniform hot-belt (b) models. For the latter we used our TDGL approach and additionally assumed that the photon energy is uniformly and instantaneously heats electrons and phonons within the area $w \times w$ and that the dynamics of the order parameter starts at $t = 0$. For the hot-spot model (Fig. 10 a) at any fixed photon energy, the difference between delay times at two different locations ($y = 5\ \xi_C$ and $y = 15\ \xi_C$) increases with the increase in magnetic field. This is also seen in Fig. 2b for the photon energy 1 eV. This corresponds to the increase in $\sigma_{loc}$ with magnetic field that we observed experimentally.

So far we did not take into account Fano fluctuations. It was argued in Refs. 6 and 33 that they are responsible for variation in that portion of the photon energy, which is delivered to the electronic system after absorption of the photon. Since the delay time depends on the photon energy delivered to electrons, this $\tau_D(h\nu)$ dependence translates Fano fluctuations into timing jitter. We further compare Fano contributions arising in frameworks of the uniform hot-belt model and the two-dimensional hot-spot model. Since for both models the fist derivative of the $\tau_D(h\nu)$ dependence is negative and its absolute value increases with the decrease in the photon energy, symmetric Gaussian PDF for Fano fluctuations will be converted into asymmetric PDF for $\tau_D$ with the tail at large delay times. Moreover, the standard deviation in the PDF for $\tau_D$ should increase with the decrease in the photon energy. Both conclusions agree with our experimental observations. For the hot-belt model, the absolute value of the first derivative decreases when the magnetic field increases. Hence, contrary to our observations, the model predicts the decrease in the jitter with the increase in the magnetic field. For the hot-spot model, the same is true when the hot-spot is located at $y = 5\ \xi_C$. At $y = 15\ \xi_C$ the tendency is exactly opposite. Jitter originating from this location should increase with the magnetic field. This qualitative analysis shows that the observed increase in the jitter in magnetic field can be understood only in the framework of the hot-spot model.

In order to distinguish between contributions to the jitter due to Fano fluctuations and the position dependence of the delay time, one need to reconstruct respective PDFs and compare them with the experiment. The reconstruction crucially depends on the local detection scenario (deterministic or probabilistic) and the uniformity of the absorbance across the strip. Generally, PDF of the random variable $z = f(x)$ which is the function of the random variable $x$ is $F(g(z))|g(z)'|$ where $F(x)$ is the PDF of the variable $x$ and $g(z)$ is the inverse function for $f(x)$. Hence, jitter due to the position-dependent delay time is entirely controlled by the function $\tau_D(y)$ shown in the inset in Fig.9a. Assuming uniform local absorbance across the wire and constant detection efficiency, we arrive to the jitter PDF with Lorentzian profile.

Jitter due to Fano fluctuations is controlled by the function $\tau_D(h\nu)$. At $B = 0$ the energy dependence of the delay time closely follows the power function $a\ (h\nu - b)^{-n} + c$ with $n = 0.75$ which is shown as dash-dotted line in Fig.10a. Assuming for the energy delivered to electrons Gaussian PDF with the standard deviation $\delta$ and the mean value $\varepsilon\ h\nu$ ($\varepsilon$ represents the quantum yield) we arrive at the jitter PDF

$$f(\tau_D) = \frac{a^{1/n}}{n\ \delta\ \sqrt{2\pi}}\ \frac{1}{(\tau_D - c)^{\frac{n+1}{n}}}\ \exp\left(-\frac{\left[\left(\frac{a}{\tau_D - c}\right)^{1/n} + b - \varepsilon\ h\nu\right]^2}{2\ \delta^2}\right). \tag{6}$$

It is worth noting that the simplest one-dimensional model for switching of the superconducting wire into the normal state under the step of supercritical current [27] predicts $\tau_D \propto I_{CD}^2$. Along with the linear current-energy relation [34], this results in the PDF of the type (6) with $n = 2$. Although the function (6) fits better our experimental PDFs than the PDF expected for the position-dependent delays, it is not capable to describe simultaneously Gaussian profile at small delays and exponential profile at large delays. To conclude, at the present stage we can not clear distinguish between contributions to the experimental jitter due to Fano fluctuations and the position-dependent delay times.

### V. CONCLUSION

We have shown that the probability density function (PDF) of timing jitter in photon detection by straight nanowire within three orders of magnitude is best described by the exponentially modified Gaussian distribution. We found experimentally that optical, instrumental and noise contributions to the measured system PDF are all obey Gaussian distribution and concluded that the exponentially modified Gaussian distribution inherent in the detection process itself. We have found that for two wavelengths 800 nm and 1560 nm both Gaussian and exponential components of the standard deviation (STD) in this intrinsic PDF monotonically increase in external magnetic field and that in the absence of the field STD is larger for the larger wavelength. Furthermore, by increasing intensity of the photon flux, we have driven the nanowire from quantum to the bolometric detection regime and have found that in the bolometric regime the exponential component of the STD disappears while the Gaussian component drastically decreases. We have shown that the difference between Gaussian components of the intrinsic STD in these two regimes scales as square root of the photon energy. Noting that such scaling is expected for Fano fluctuations we associate this difference with fluctuation in the portion of the photon energy which is delivered to the electrons.

We accounted for magnetic field in the framework of the two-dimensional deterministic hot-spot detection model and shown that the hot-spot is essential in explaining magnetic field dependence of the intrinsic STD: magnetic field dependence predicted by the uniform hot-belt detection model disagrees with our experimental observations. Although being capable to explain the effect of magnetic field, the hot-spot model along with Fano fluctuations does not reproduce experimental intrinsic PDF.

## ACKNOWLEDGEMENT

M.S. acknowledges support by The Helmholtz Research School on Security Technologies (HRSST), D. Yu. V. acknowledges support from the Russian Scientific Foundation (RSF), grant No. 17-72-30036.

## APPENDIX:

### A. Optical jitter

We estimate optical contribution to the measured jitter from the known fiber parameters and the pulse spectral width and also evaluate it experimentally for two fiber types (SMF28 and SM980) at the wavelengths 800 nm and 1560 nm. Standard deviation of the photon flight-time through the fiber $\sigma_{opt} = |D|\sigma_\lambda L$ is proportional to the fiber length, $L$, spectral width $\sigma_\lambda$ of the optical pulse, and the dispersion coefficient $D$. At the wavelength 800 nm spectrum of the laser pulse is Fourier-transform limited and has Gaussian profile with the STD $\sigma_\lambda$ = 19 nm. For both studied fibers at this wavelength, physical dispersion provides the dispersion coefficient $D$ = -120 ps/(km×nm). Hence, due to physical dispersion alone at 800 nm we expect optical jitter per unit length of the fiber 2.3 ps/m. At the wavelength 1560 nm, the spectrum of the pulse can be modelled as a sum of several Gaussian lines. Overall spectral STD is given by the square root of the sum of the squares of particular STD of each Gaussian line and amounts at $\sigma_\lambda$ = 28.6 nm. For single-mode fiber SMF28, the physical dispersion at 1560 nm provides the coefficient $D$ = 18 ps/(km×nm) that results in 0.5 ps/m.

To evaluate optical contribution experimentally, we carried out several measurements with different length of the fiber between the light source and the nanowire [8]. At this stage we suppose that the flight time of the photon and the time delay in the detection of the photon are not correlated. We prove that statement in the next section. In this case the measured variance in the total delay time of the appearance of the photon count is the sum of the variance in the flight-time of the photon and the variance, $var$, in the remaining part of the total delay time. The measured system jitter takes the form $\sigma_{sys} = (var + (\sum_{i=0}^{N} \sigma_i)^2)^{1/2}$ where $\sigma_i$ – the contribution to the jitter added by each of $N$ pieces of the fiber with its particular length. Each combination of fibers provides an independent equation of this type. Since the operation regime of the nanowire and the electronics remain unchanged during measurements, the variance $var$ was constant. Solving the system of several linear equations for different fiber combinations, we found optical jitter per unit length 0.5±0.1 ps/m in SMF28 at 1560 nm. At 800 nm the jitter per unit length was 2.4±0.5 ps/m and 3.9±0.5 ps/m in SM980 and in SMF28, respectively. Therefore, for the combinations used in the experiment, optical contribution was $\sigma_{opt}$ = 3 ps at 1560 nm (6 m of SMF28) and $\sigma_{opt}$ = 8 ps at 800 nm (2 m of SMF28 plus 2 m of SM980). For the wavelengths 1560 nm in SMF28 and for the wavelength 800 nm in SM980, the measured optical jitter per unit length agrees well with the jitter per unit length estimated for the physical dispersion alone. For the wavelength 800 nm in SMF28, the measured value is larger. This is because at 800 nm the fiber operates at the border of single-mode regime.

As a reference, at 800 nm we evaluated optical contribution of the experimental combination of fiber in a different way. We measured system jitter for the same set of bias currents with and without fibers and obtained a set of equations in the form $\sigma_{sys}(I) = (var(I) + \sigma_{opt}^2)^{1/2}$. We found an average value of the optical jitter 8±0.3 ps that agrees well with the result obtained by the other method.

### B. Probability density function of the system jitter.

The unique feature of PDFs which we obtained experimentally is their exponential profile at large delay times and Gaussian profile at small delay times. The both profiles are preserved down to a level of $10^{-3}$ from the maximum. This is shown in Fig. B1. As it is shortly discussed in the main text (Sec. IV B), we describe experimental PDFs as a conditional joint PDF of the delay time produced by a series of sequential events. The joint PDF is a multiple integral of the form [21]

$$F(t) = \int_{-\infty}^{\infty} \left\{ \int_{-\infty}^{t_4} \left[ \int_{-\infty}^{t_3} \left( \int_{-\infty}^{t_2} f_1(t_1) f_2(t_2 - t_1) dt_1 \right) f_3(t_3 - t_2) dt_2 \right] f_4(t_4 - t_3) dt_3 \right\} f_5(t - t_4) dt_4, \qquad \text{(B1)}$$

where $f_i(t)$ is the PDF of the delay time associated with the $i$-th particular event. The total delay-time in each realization (photon count) is the sum of particular delays which are considered as independent random variables.

Consequently, the mean delay is the sum of mean values for particular delays. In Eq. (B1) we include five particular events with their corresponding variables: travelling time of the photon in the fiber ($i = 1$), delay in the emergence of the hot-spot ($i = 2$), delay in the start of vortex crossing ($i = 3$), travelling time of the signal from the hot-spot to the amplifier ($i = 4$, geometric jitter), joint contribution of electrical noise and instruments ($i = 5$). All these variables are statistically independent, however, for $i = 1 - 4$, variables affiliated with adjacent events are connected via conditional probability, i.e. the later event occurs only if and after the earlier has occurred. Conditional probability also implies that, for instance, $f_2(t_2) = 0$ at $t_2 < 0$. The effective delay added by noise and instruments is not connected with the previous delay by conditional probability. Consequently, the joint PDF is defined by the first integral from the left where the upper integration limit is set to plus infinity. Joint PDF can be evaluated either via sequential integration or by combining adjacent particular PDFs.

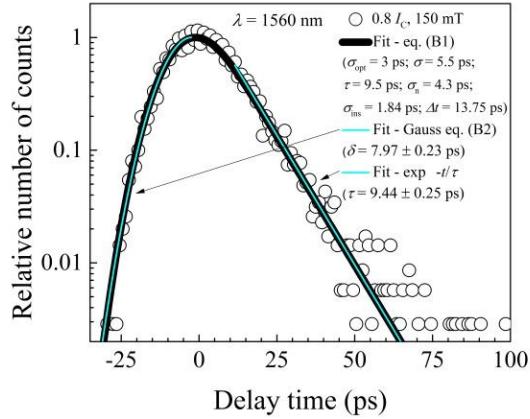

FIG. B1. Exemplary PDF obtained for the wavelength 1560 nm at the bias current 0.8 $I_C$. Thick line shows the best fit with eq. (B5). Fit parameters are specified in the legend. Thin line represents Gaussian function $f(t, \delta)$ (B2) with $\delta = (\sigma_{\text{opt}}^2 + \sigma^2 + \sigma_n^2 + \sigma_{\text{ins}}^2)^{1/2}$. Thin straight line is the function $\exp(-t/\tau)$.

We suppose that for Gaussian spectra of laser pulses and relatively short fiber lengths PDF of the optical delay (flight-time of the photon trough the fiber) $f_1(t)$ has Gaussian form [35]

$$f(t, \delta) = \frac{1}{\delta\sqrt{2\pi}} \exp\left(-\frac{t^2}{2\delta^2}\right). \tag{B2}$$

with the standard deviation $\delta$ equal to $\sigma_{\text{opt}}$ (Appendix A)

Following the approach of the vortex-assisted photon counts [1, 16, 25 7], we suppose that PDF $f_2(t)$ associated with the growth time of the hot-spot represents Gaussian distribution (A2) with the standard deviation $\sigma$, while PDF of the time-delay between emergence of the hot-spot and the start of the vortex-crossing is characterized by the standard deviation $\tau$ and has exponential form $f_3(t) = \frac{1}{\tau}\exp\left(-\frac{t}{\tau}\right)$. Combining these two sequential PDFs, we obtain PDF for the time delay inherent in the detection process

$$g(t, \sigma, \tau) = \int_{-\infty}^{t} f(u, \sigma) f_3(t - u) du = \frac{1}{2\tau} \exp\left(\frac{1}{2\tau}\left(\frac{\sigma^2}{\tau} - 2t\right)\right) \cdot \left(1 - \text{erf}\left(\frac{\frac{\sigma^2}{\tau} - t}{\sigma\sqrt{2}}\right)\right). \tag{B3}$$

This is well known exponentially modified Gaussian distribution. With a cumbersome math [36] it can be shown analytically that the mean value and standard deviation for this distribution are $\tau$ and $(\sigma^2 + \tau^2)^{1/2}$, correspondingly. This distribution has Gaussian shape for $t \leq t_0$ and exhibits exponential tail at $t > t_0$ where $t_0$ maximizes $g(t, \sigma, \tau)$. These approximations are shown in Fig. B1. Although possible, combining analytically $f_1(t)$ and $g(t)$ represents a serious challenge.

Alternatively, the integral can be quickly solved by following the natural sequence of events. Combining Gaussian PDFs $f_1(t)$ and $f_2(t)$ (round brackets in B1) in the form of (B2), we find [36]

$$p(t) = \int_{-\infty}^{t} f(u, \sigma_{\text{opt}}) f(t - \Delta t - u, \sigma) du = \frac{1}{2\sqrt{2\pi(\sigma_{\text{opt}}^2 + \sigma^2)}} \exp\left[-\frac{(t-\Delta t)^2}{2(\sigma_{\text{opt}}^2 + \sigma^2)}\right] \cdot \left(1 + \text{erf}\left[\frac{(t-\Delta t)}{\sqrt{2(\sigma_{\text{opt}}^2 + \sigma^2)}}\frac{\sigma}{\sigma_{\text{opt}}} + \Delta t \frac{\sqrt{\sigma_{\text{opt}}^2 + \sigma^2}}{\sigma_{\text{opt}} \sigma \sqrt{2}}\right]\right)$$

(B4)

where $\Delta t$ is the delay-time between photon absorption and emergence of the hot-spot. For $\Delta t \geq 2.5\,\sigma$, the joint PDF, $p(t)$, is almost Gaussian and its standard deviation equals $(\sigma_{\text{opt}}^2 + \sigma^2)^{1/2}$. For smaller $\Delta t$, $p(t)$ quickly becomes asymmetric and non-Gaussian. Experimentally, the delay time $\Delta t$ is associated with the rise-time of the voltage response in the bolometric regime. For NbN the rise-time of the order of 10 ps was reported [37]. In our microscopic hot-spot model $\Delta t$ is associated with the time-delay $\tau_D$. Its smallest value (Fig. 10a) is approximately $15\tau_C$. For our nanowires $\tau_C \approx 0.6$ ps that gives the time delay of approximately 10 ps. In the fitting procedure we used Gaussian form for $p(t)$, i.e. we assumed $p(t) = f(t, (\sigma_{\text{opt}}^2 + \sigma^2)^{1/2})$. Although best fit values for $\sigma$ are comparable to the expected $\Delta t$ values, deviations from the Gaussian shape in experimental PDFs at small delays are below the accuracy of our measurements. This observation justifies the self-consistency of our fitting procedure.

Further combining $p(t)$ in its Gaussian form with $f_3(t)$, we obtained exponentially modified Gaussian distribution in the form of $g\left(t, \sqrt{\sigma_{\text{opt}}^2 + \sigma^2}, \tau\right)$ with the standard deviation $(\sigma_{\text{opt}}^2 + \sigma^2 + \tau^2)^{1/2}$. Here we neglect geometrical contribution to the jitter ($f_4$ in (B1)) which is below the accuracy of our measurements (see Sec. IV A). Finally, we add noise and instruments and obtain joint PDF in the analytical form [38]

$$F(t) = \int_{-\infty}^{\infty} g\left(u, \sqrt{\sigma_{\text{opt}}^2 + \sigma^2}, \tau\right) f\left(u - t, \sqrt{\sigma_n^2 + \sigma_{\text{ins}}^2}\right) du = \frac{1}{2\tau} \exp\left(\frac{1}{2\tau}\left(\frac{\sigma_*^2}{\tau} - 2t\right)\right) \cdot \left(1 - \text{erf}\left(\frac{\frac{\sigma_*^2}{\tau} - t}{\sigma_* \sqrt{2}}\right)\right) \quad \text{(B5)}$$

with $\sigma_* = (\sigma_{\text{opt}}^2 + \sigma^2 + \sigma_n^2 + \sigma_{\text{ins}}^2)^{1/2}$. Function $F(t)$ represents exponentially modified Gaussian distribution which we used to fit our experimental PDFs and to evaluate components of the system jitter. The system jitter which is the standard deviation in $F(t)$ is given as

$$\sigma_{\text{sys}} = (\sigma_{\text{opt}}^2 + \sigma^2 + \tau^2 + \sigma_n^2 + \sigma_{\text{ins}}^2)^{1/2}. \quad \text{(B6)}$$

In our fitting procedure we used only two free parameters $\sigma$ and $\tau$, while other parameters ($\sigma_{\text{opt}}$, $\sigma_n$, $\sigma_{\text{ins}}$) were set at the values independently defined from measurements. The best fit of an exemplary PDF with $F(t)$ is shown in Fig. B1.

_______________________________________